# Einstein's 1927 gedanken experiment: how to complete it and measure the collapse time of a spatially spread photon.

Alejandro A. Hnilo
*CEILAP, Centro de Investigaciones en Láseres y Aplicaciones, UNIDEF (MINDEF-CONICET);*
*CITEDEF, J.B. de La Salle 4397, (1603) Villa Martelli, Argentina.*
*email: alex.hnilo@gmail.com*

In the famous Solvay 1927 conference, Einstein discussed a gedanken experiment involving a single photon diffracted at an aperture and impinging on a screen. He devised the example to support De Broglie's hypothesis of the pilot wave, and his own ideas on the incompleteness of the description of physical reality provided by Quantum Mechanics. Partial realizations of Einstein's example have been performed, but the complete experiment has not been attempted (for good practical reasons) yet. Here I describe how to do it with accessible means, by inserting a transmission grating. Among other possibilities, the setup will allow testing Hellwig and Kraus' postulate of covariant reduction.

When Quantum Mechanics (QM) was still maturing, A.Einstein proposed the following situation, that simultaneously mixed the particle and wave features of a quantum state: consider a *single* photon that, after being diffracted at an aperture, is detected on a distant screen (Figure 1). There is no question the photon is detected in only one point on the screen; otherwise energy would not be conserved. However, he was concerned about the situation in which detectors are space-like separated, making impossible any "coordination" between them to decide which one would fire. As it is quoted in [1]:

*"It seems to me that this difficulty cannot be overcome unless the description of the process in terms of the Schrödinger wave is supplemented by some detailed specification of the localization of the particle during its propagation. I think M. de Broglie is right in searching in this direction."*

De Broglie's pilot wave would "guide" the single photon after the aperture. The photon would follow a well defined trajectory to the screen (where it is eventually detected), while the pilot wave would reproduce the diffraction pattern for a large number of photons. This explanation "supplemented" QM with elements that are not in the standard theory. Einstein later refined his ideas on the incompleteness of QM description of physical reality in the famous "EPR paradox". De Broglie's ideas were later elaborated by D.Bohm to accurately reproduce QM predictions (Bohmian Mechanics).

An alternative explanation is to hypothesize energy is conserved in the average, and not "photon per photon". This would allow two photons to be detected on the screen sometimes, what would be compensated by other times in which none is detected. This alternative is consistent with the theory of stochastic quantum dynamics [2]. It is pertinent to say here that "energy conserved in the average" is the correct result if the experiment in Fig.1 is performed with classical light. For a coherent quantum state (the closest to classical light) the number of photons is not well defined. For a "one photon" (in the average) coherent state, standard QM predicts that there is in fact a nonzero probability that the two detectors fire simultaneously, and a nonzero probability that none fires. In order to realize Einstein's example, the state of the field must be non classical (a single photon Fock state). The details of this situation according to standard QM have been thoroughly discussed in [3].

A first test of Einstein's example was performed with the then newly developed atomic single photon sources by A.Aspect's group in Orsay [4], verifying that energy was conserved "photon per photon". This result was later confirmed in improved setups, with space-like separated detectors [1] and then by measuring the time required by the photon state to collapse, establishing a lower bound for the velocity of the hypothetical interaction "coordinating" the detectors. This bound was $\geq 1550c$ (where $c$ is the velocity of light) [3].

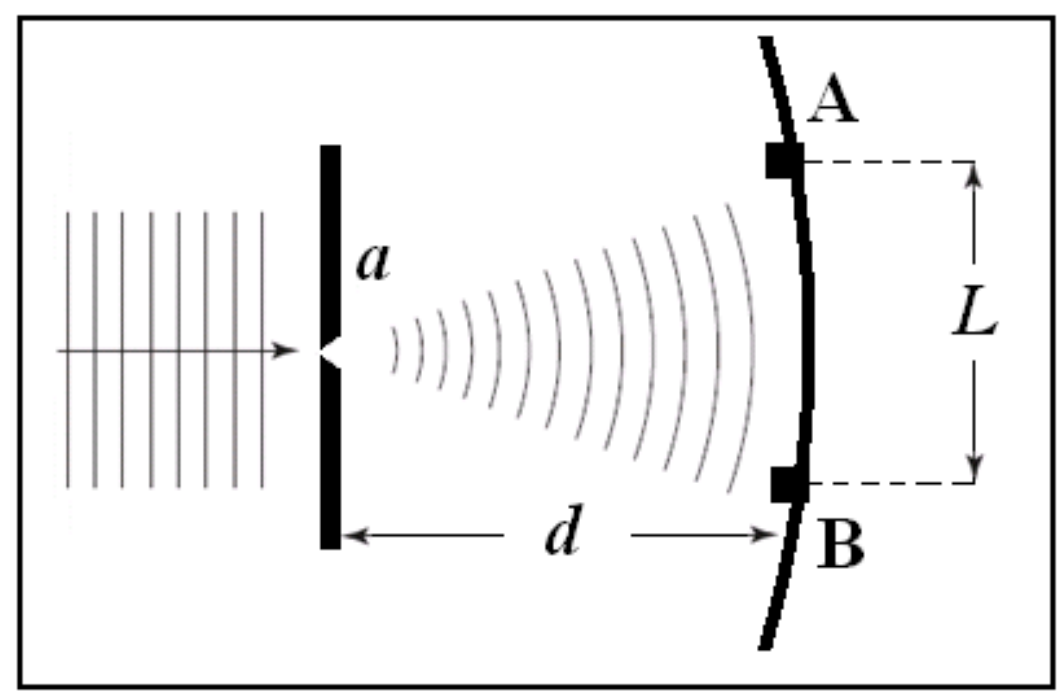

Figure 1: Einstein's 1927 gedanken experiment. A *single* photon is diffracted at an aperture and two detectors are placed distant on a screen, such that the time $L/c$ is much longer than the time required to detect a photon. Only one detector fires, energy would not be conserved otherwise. The question is how detectors "coordinate" which one fires, if there is no information additional to the wavefunction (i.e., if QM is complete). This example was presented by Einstein to support De Broglie's idea of the pilot wave guiding the photon's trajectory, thus, that QM had to be "supplemented". This example was somehow left behind after the discussion on the 1935 "EPR paradox" and Bell's inequalities.

Nevertheless, all these experiments used a beam splitter to direct the photon to detectors A and B. There are good practical reasons for that (see later), but the beam splitter erases the "wave" feature of diffracted propagation; the photon might have chosen its direction of propagation at the beam splitter. Much of the appeal of Einstein's original example is that the photon behaves both as wave (because of diffraction) and as particle (because of conservation of energy).

In this paper I describe how to perform Einstein's gedanken experiment as it was originally stated. Such a setup will be useful to explore several features of QM. In particular, it will allow testing whether the collapse (or reduction, or projection) of the quantum state is instantaneous, as stated in the standard Copenhagen interpretation of QM, or not.

The value of the collapse time is related with the discussion on the nature of the quantum state, that is: whether it is physically real (*ontic* nature) or if it only represents the information the observer has about the system (*epistemic* nature) [5]. If the first alternative is correct, then it is natural to expect the collapse time to take a finite value [8]. The second one, instead, means the collapse to correspond to a change of the information accessible to the observer. It occurs in the observer's mind (as J. von Neumann argued), it is therefore not physical, and can be instantaneous. The second alternative is the one assumed in standard QM. The drawback of this alternative is that it leaves the door open to the possibility that part of reality is left unaccounted by the quantum state (as Einstein argued). The epistemic alternative has been claimed refuted [6], but this refutation has been recently contested [7].

In summary: revealing the nature of the quantum state is an important open problem. Testing the value of the collapse time may put some light on this problem. Proposals to test the collapse time, as it is predicted by existing theories of collapse dynamics, have been reported [3,8-10]. In particular, Hellwig and Kraus postulate of covariant state reduction [11,12] implies the collapse time to be given by the state's spatial spread scaled with *c*. This postulate is attractive, for it offers an elegant explanation to "non locality" when Bell's inequalities are violated, still allowing the results to be fully consistent with Relativity. In the case of Fig.1, it means the photon detection time is delayed by a time value that depends of the diffraction spread on the screen. Numbers explain why this effect, even if it exists, has not been observed before (see below).

First, let consider the original setup in Fig.1, and why beam-splitters have been used. In order to avoid any "coordination" between the detectors, the time resolution to reliably detect a photon (≈2 ns for an avalanche photodiode) must be much shorter than $L/c$. Assuming $L/c$ > 10 ns, then $L$ > 3m. This forces the diffracted beam to spread over a large area, reducing the intensity to an unpractical level. Let see the numbers: the single photon state impinging on the aperture from the left is created by detecting one photon of an entangled pair in some other detector. In avalanche photodiodes, a valid detection has some probability of triggering spurious "echo" detections. If the number of spurious detections is large, the observation is jammed. The echoes' probability increases when the detections' rate comes close to $\approx 10^5$ $s^{-1}$, which is therefore the upper limit for the rate of creation of reliable single photon states. Because of imperfect alignment, transmission losses and losses at the diffracting aperture, the rate of photons impinging on the screen is lower. Let suppose a good efficiency of 25%. Finally, assume detectors or collecting optics at the screen to have a (typical) diameter of 5mm, and 70% efficiency. This means the rate of photons detected at the screen is smaller than $10^5 \times 0.25 \times (5\text{mm}/3\text{m}) \times 0.7\ s^{-1} < 30\ s^{-1}$, which is below the typical rate of dark counts (or noise, $\approx 100\ s^{-1}$) of avalanche photodiodes with 70% efficiency. That's why the experiment as in Fig.1 is impossible nowadays. That is also the reason why beam splitters have been used to send the photons directly to the detectors in the performed experiments. A beam splitter improves the estimated number of detected photons a factor $>10^3$, making the observation of the collapse time feasible.

The proposal here is to use a diffraction grating instead of a single diffracting aperture. The photon is, of course, diffracted following the same physical principles of wave propagation as in Fig.1, but losses are smaller and intensity distribution on the screen more favorable. A sketch of the setup is shown in Figure 2. In the Fraunhofer approximation, the distribution of intensity I($x$) on the screen is given by:

$$I(x) \propto (\sin^2(\beta)/\beta^2).\sin^2(N.\gamma)/\sin^2(\gamma) \qquad (1)$$

where $\beta = (\pi a/\lambda).\sin(\theta)$, being $a$ the width of each transmitting aperture in the grating and $\lambda$ the wavelength; $\gamma = (\pi p/\lambda).\sin(\theta)$, being $p$ the grating's period; $N$ is the (usually large) number of illuminated apertures. Finally, $x/f = \theta$.

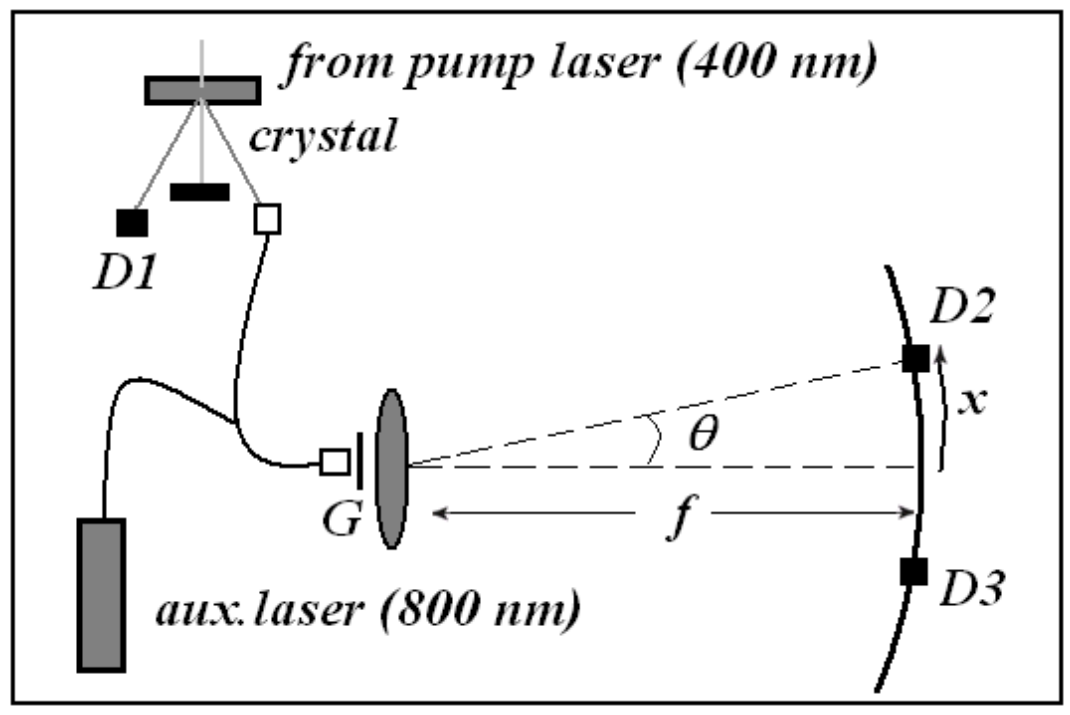

Figure 2: Sketch of the proposed setup. A CW pump laser (out of the figure), a nonlinear crystal and detector *D1* prepare the single photon state. Open squares indicate focusing optics into and out of a (multimode) optical fiber; G is a transmission grating. Diffracted light is focused to a screen where single photon detector *D2* (*D3*, etc.) is placed at position *x*. A time-to-digital converter (not indicated) records the time values photons are detected for further processing. An auxiliary fiber-coupled laser is used for alignment; it is removed during photon counting (see the main text).

Let suppose a symmetric ($p$=2$a$) transmission grating 800 lines/mm, λ=800 nm and that ≈3mm of the grating are illuminated ($N$>2000). In such situation, I($x$) is concentrated in three sharp peaks, see Figure 3. The central, main one has maximum intensity $I(0) \propto N^2$, the two lateral ones (corresponding to $\gamma = \pm\pi$) have maximum intensity $\propto 4.\mathrm{I}(0)/\pi^2$. Therefore, each lateral peak has ≈20% of the total intensity diffracted on the screen. Using the same numbers as before, detectors placed at the secondary peaks will deliver $\approx 2\times10^4$ $s^{-1}$ counts each, well above dark noise. Einstein's original gedanken experiment can be realized in this way with accessible means.

The proposed setup is able to observe, f.ex., the time delay of detection according to Hellwig and Kraus theory of covariant state reduction (or collapse). In this theory, it is postulated that a quantum state collapses along the past light cone of the observation event. If the state is originally spread over a distance $s$, then the time required for its collapse is $s/c$. Therefore, a state with a larger spread $S$ is observed a time $(S\text{-}s)/c$ after a state with spread $s$. The idea is simple, but there is a problem in the definition of "spatial spread". F.ex., a state with Gaussian profile $exp(-x^2/\sigma^2)$ is, in principle, infinitely spread and hence should be observed after an infinite time. However, it is intuitive that the "effective" spread must be somehow proportional to σ. In other words, there must be some sort of weighed contribution, proportional to the intensity in each point, to define the spread's value. I propose:

$$\int dx.\,|x|\,.\mathrm{I}(x)\,/\int dx.\mathrm{I}(x) \equiv spread \qquad (2)$$

where $x$ is the distance measured from the center of mass of $\mathrm{I}(x)$. Then $c.\mathrm{T}_{delay} = spread$, where $\mathrm{T}_{delay}$ is the time elapsed since the moment of detection of a narrow reference distribution $\mathrm{I}(x)_{reference}$ ("narrow" means spatial width smaller than time resolution × $c$) and centered in $x$=0, say $\mathrm{I}(x)_{reference} \approx \mathrm{I}_0.\delta(x)$.

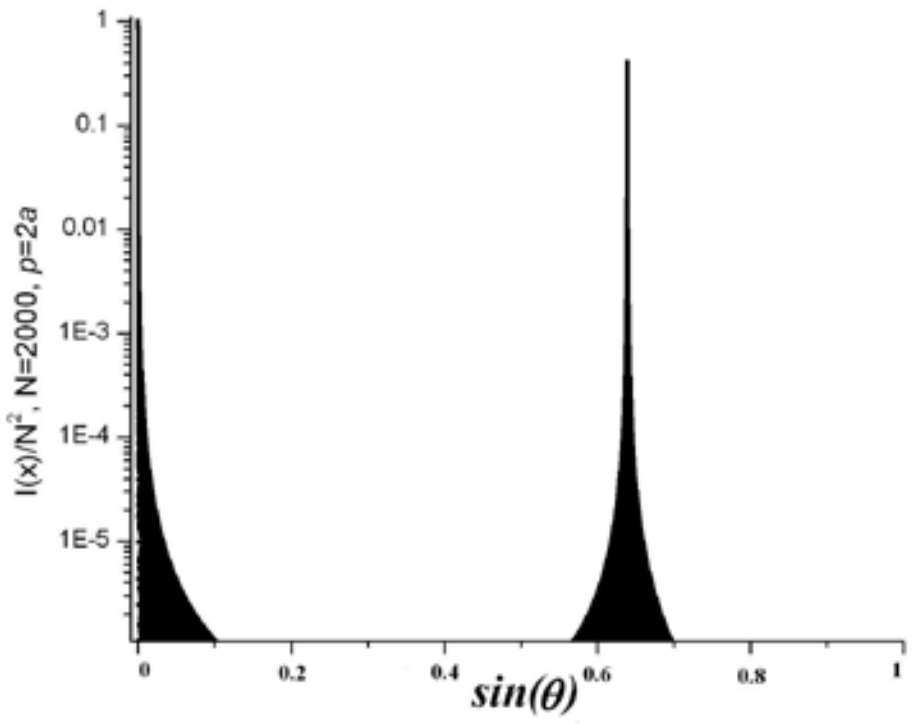

Figure 3: Normalized $\mathrm{I}(x)$ as a function of $sin(\theta)$ for the setup described in the text, for positive values of θ. A third peak is symmetrically placed at θ = -0.694. Note the logarithmic vertical scale.

F.ex., let suppose that the intensity on the screen consists of two identical narrow peaks separated by a distance $q$, then: $\mathrm{I}(x) = \frac{1}{2}.\mathrm{I}_0.[\delta(x\text{-}q/2) + \delta(x+q/2)]$. Inserting this $\mathrm{I}(x)$ in eq.2, then $\mathrm{T}_{delay} = q/2c$. This is the time value detections would be observed after the ones produced, in the same conditions, by the distribution $\mathrm{I}_0.\delta(x)$.

For the grating supposed before, $\mathrm{I}(x)$ is symmetrical with respect to the central peak, and the relative intensities of the three peaks is independent of $N$. Then, from eq.2: $c.\mathrm{T}_{delay} \approx (0.\mathrm{I}_0+2.q.\mathrm{I}_0.4/\pi^2)/\ (\mathrm{I}_0+ \mathrm{I}_0.8/\pi^2) \approx 0.448\times q$, where $q$ is the position of the lateral peaks (Fig.3). This position is given by the condition $(\pi p/\lambda).sin(\theta) = \pm\pi$, then $\theta \approx 0.694$, and $q = f.\theta$. Assuming $f$ = 4m:

$$\mathrm{T}_{delay} \approx 0.448\times0.694\times4\mathrm{m}/c \approx 4 \text{ ns} \qquad (3)$$

It is pertinent asking why this delay effect has not been reported before, if it existed. In the first place, it is a small value, barely above the time resolution of single photon detectors. Secondly, recall the diffracted state must be a *single* photon one (i.e., a non-classical state of the field). For classical light $\mathrm{T}_{delay}$ is zero. It is the conservation of energy implicit in the single photon state what "compels" the delay. Finally, note the unusual intensity distribution: in the proposed setup, the lateral peaks on the screen are at ≈2.5 m (≈8 ns) from the central one. The effect is so small, that it can be discerned only if it is intentionally looked for in a specifically designed experimental setup.

The observation should run as follows (see Fig.2). An auxiliary laser at 800 nm is inserted in the optical fiber to facilitate focusing and alignment. If the grating G is removed, the intensity on the screen is a narrow spot, or $\mathrm{I}(x)_{reference} \approx \mathrm{I}_0.\delta(x)$. Place detector *D2* at this spot (what defines $x$=0, or θ=0 in an ideal setup), then remove the auxiliary laser and insert the fiber from the source of single photons. Series of time values of photons' detections at *D1* and *D2* are stored in a single time-to-digital converter until gathering sufficient statistics. The series are processed in the usual way to find the delay value $\mathrm{T}_0$ that produces the maximum of coincidences between *D1* and *D2*. The coincidence time window should be set to the detectors' time resolution, ≈2 ns. The value of $\mathrm{T}_0$ is the time difference between detections at *D1* and *D2* due to different cable and fiber lengths, free space propagation and detectors' response times.

Then insert the grating and repeat the observation, obtaining a delay value $\mathrm{T}_1$ for maximum of coincidences between *D1* and *D2*. If the postulate of covariant reduction is correct, from eq.3:

$$\mathrm{T}_1 - \mathrm{T}_0 \geq \mathrm{T}_{delay} \approx 4 \text{ ns} \qquad (4)$$

According to standard QM, $\mathrm{T}_1 - \mathrm{T}_0 = 0$ instead. The difference between the two predictions can be reliably established by gathering sufficient statistics, even with only 2 ns time resolution. Besides, be aware that $\mathrm{T}_{delay}$ is the *lower bound* of $\mathrm{T}_1 - \mathrm{T}_0$.

In summary: a setup able to reproduce the original Einstein's gendanken experiment of 1927 is described. It is achievable with means available in almost all quantum optical labs nowadays. The setup would be able to explore the features of Einstein's example at will. Among the possibilities to explore, it would allow to directly test Hellwig and Kraus postulate of covariant state reduction. Observing the result in eq.4 would give decisive support to the idea of the ontic nature of the quantum state, and would deeply affect the foundations of QM as they are currently known. On the other hand, observing $T_1 - T_0 \approx 0$ would be consistent with the epistemic nature of the quantum state and the standard interpretation of QM, but this result would be relevant too. As it is recently stated [13]: *"So it is fair to celebrate 2025 as the true centenary of quantum theory. Although such a commemoration can rightly point to a wide variety of breathtaking experimental successes, it must leave room to acknowledge the foundational questions that remain unanswered. Quantum mechanics is a beautiful castle, and it would be nice to be reassured that it is not built on sand".*

**Acknowledgments.**

This contribution received support from grant PIP 00484-22 from CONICET (Argentina).

**References.**